\documentclass[12pt]{article}
\usepackage{graphicx}

\newcommand\pubnumber{}
\newcommand\pubdate{\today}

\def\napoli{Physikalisches Insitut\\
Heidelberg University, Philosophenweg 12, 69120 Heidelberg, GERMANY}

\def\Title#1{\begin{center} {\Large #1 } \end{center}}
\def\Author#1{\begin{center}{ \sc #1} \end{center}}
\def\Address#1{\begin{center}{ \it #1} \end{center}}

\newcommand\pubblock{\rightline{\begin{tabular}{l} \pubnumber\\
         \pubdate  \end{tabular}}}
\newenvironment{Abstract}{\begin{quotation}  }{\end{quotation}}
\newenvironment{Presented}{\begin{quotation} \begin{center} 
             PRESENTED AT\end{center}\bigskip 
      \begin{center}\begin{large}}{\end{large}\end{center} \end{quotation}}

\begin{document}
\begin{titlepage}
\pubblock

\vfill
\Title{Measurement of $\phi_s$ at LHCb}
\vfill
\Author{ Stephanie Hansmann-Menzemer}
\Address{\napoli}
\vfill
\begin{Abstract}
A time dependent angular analysis of the decay mode $B_s \rightarrow J/\psi
\phi$ allows for the measurement of the mixing induced CP-violating phase
$\phi_s$. Within the Standard Model $\phi_s$ is theoretically precisely
predicted to be very small, however many Standard Model extensions predict sizeable
contributions to this phase \cite{nierste}. The current experimental knowledge  of $\phi_s$
has very larger uncertainties. However already with the data expected to be
delivered within the next year, the LHCb experiment at the Large Hadron Collider at CERN,
has the potential to improve significantly existing measurements.\\
In a data set of up to 37.5 pb$^{-1}$ taken in 2010, first physics signals in the LHCb detector
are reconstructed and their properties are compared to Monte Carlo predictions.
Based on recently published measurements of $b\bar{b}$ cross-sections from the
LHCb collaboration \cite{sheldon}, the sensitivity on the $CP$ violating phase $\phi_s$ in
the decay $B_s \rightarrow J/\psi \phi$ is evaluated. \\
Additionally an alternative method to potentially extract complementary information 
on  $\phi_s$ from the measurement of the asymmetry in semileptonic final states is presented.

\end{Abstract}
\vfill
\begin{Presented}
6$^{th}$ International Workshop on the CKM Unitarity Triangle\\
Warwick, United Kingdom,  September 6-10, 2010
\end{Presented}
\vfill
\end{titlepage}
\def\thefootnote{\fnsymbol{footnote}}
\setcounter{footnote}{0}

%%%%%%%%%%%%%%%%%%%%%%%%%%%%%%%%%%%%%%%%%%%%%%%%%%%%%%%%%%%%%%%%%%%%%%%%%%%
%

\section{Introduction}
The phenomenological aspects linked to $B_s \rightarrow J/\psi \phi$ decays are
discussed in many articles \cite{angularanalysis}. The main parameters 
involved are introduced briefly here.\\
$|B_s>$ and $|\bar{B_s}>$ are flavour eigenstates with the quark
content: $|\bar{b}s>$ and $|b\bar{s}>$ respectively. Any arbitrary combination
of flavour eigenstates has a time evolution described by an effective Schr\"odinger equation:

\begin{center}
$i\frac{\mathcal{\partial}}{\mathcal{\partial} t}\left( \begin{array}{c} |B_s(t)>
    \\ |\overline{B_s(t)}> \end{array} \right)
 = \left( \mathrm   \bf{M^s - \frac{i}{2}
    \Gamma^s} \right)
\left( \begin{array}{c} |B_s(t)> \\ |\overline{B_s(t})> \end{array} \right)$\\
\end{center}
where {\bf M} and {\bf \boldmath $\Gamma$} are $2\times2$ hermitian matrices.
The heavy and light mass eigenstates of the Schr\"odinger equation are given
via:
\begin{eqnarray} 
|B_H> &=& p|B_s> - q|\bar{B_s}>, \nonumber \\
|B_L> &=& p|B_s> + q|\bar{B_s}>. \nonumber 
\end{eqnarray}
The complex coefficients $p$ and $q$ obey the normalizations condition: $|p|^2
+ |q|^2 =1$. 
The mass difference $\Delta m_s$ and the width difference $\Delta \Gamma_s$
between the mass eigenstates are defined by:
\begin{center}
$\Delta m_s = M_H - M_L$,~~~~$\Delta \Gamma_s = \Gamma_L - \Gamma_H$.
\end{center}
Hence, the average mass and width can be written:
\begin{center}
$M_{B_s} = \frac{M_H + M_L}{2}$,~~~~$\Gamma_s = \frac{\Gamma_L + \Gamma_H}{2}$
\end{center}
$\Delta m_s$ and $\Delta \Gamma$ are related to the Hamiltonian of the
Schr\"odinger equation via  $\Delta m_s = 2|M_{12}|$ and $\Delta \Gamma =
2|\Gamma_{12}|\cos \phi_s$ respectively.
The phase is defined as $\phi_s=arg(-\frac{M_{12}^s}{\Gamma_{12}^s})$ and
predicted to be $(3.40^{+1.32}_{-0.77}) \times 10^{-3}$ rad \cite{nierste} within the Standard
Model.\\

\begin{figure}[htb]
\begin{minipage}{0.33\textwidth}
\centering
\includegraphics[height=2.3cm]{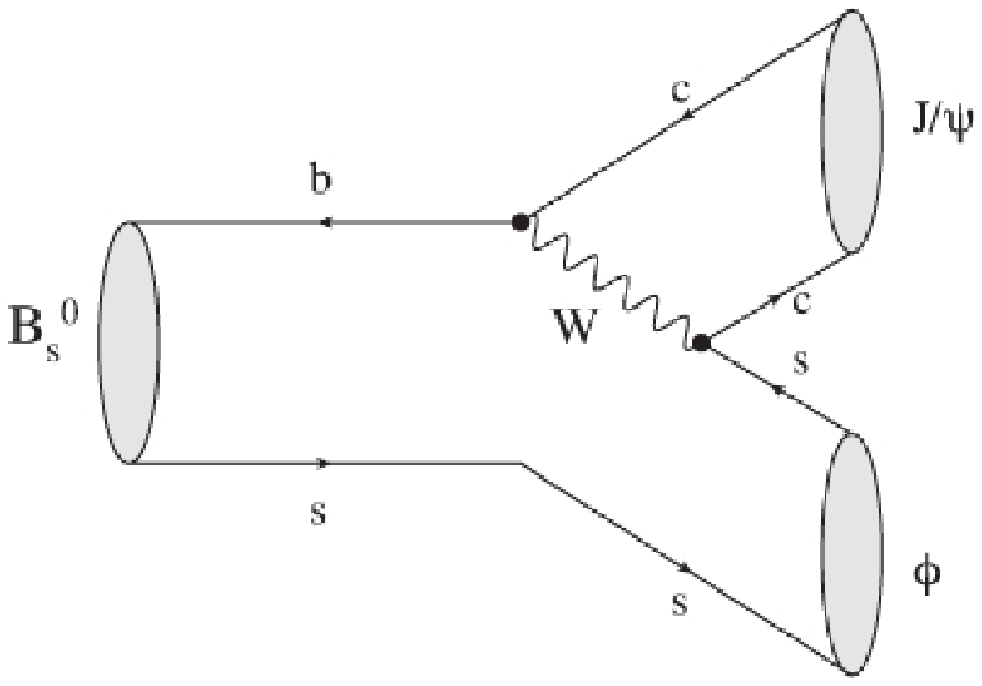}
\caption{Tree-level decay of $B_s \rightarrow J/\psi \phi$.}
\label{fig:SHM_1}
\end{minipage}
\begin{minipage}{0.67\textwidth}
\centering
\includegraphics[height=2.3cm]{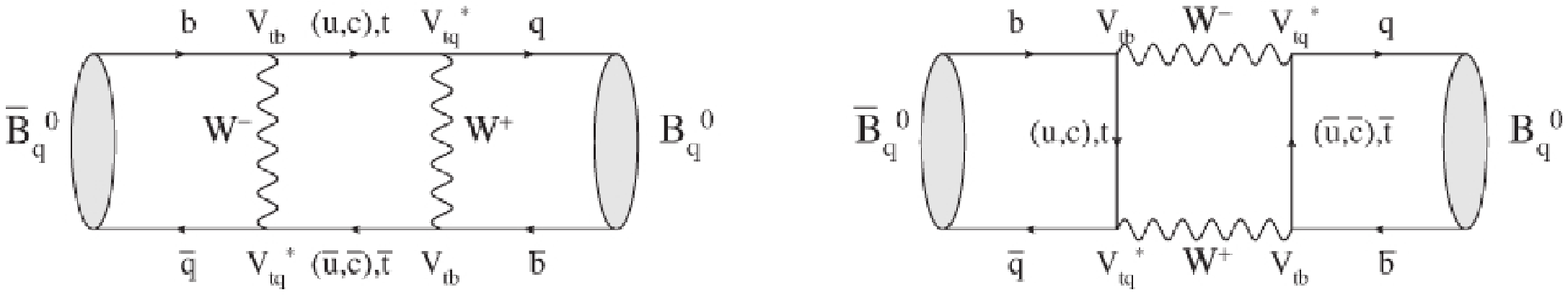}
\caption{$B_q -\bar{B_q}$ mixing diagrams (q=s,d).}
\label{fig:SHM_2}
\end{minipage}
\end{figure}

The decay $B_s \rightarrow J/\psi \phi$ is dominated by a single tree level
decay with the complex phase $\phi_D$ (Fig. \ref{fig:SHM_1}).
The $B_s$ can either directly decay into the final state $J/\psi \phi$ or
first mix into its antiparticle $\bar{B_s}$ via the box diagram displayed in
Figure \ref{fig:SHM_2}, and then decay into the same final state.
The observable phase $\phi^{J/\psi \phi}$ which we will measure in the presented analysis is the
phase difference of the phase of the mixing diagram $\phi_M$ and th phase of the decay: $\phi_s^{J/\psi \phi} = \phi_M - 2\phi_D$.
Neglecting any contributions from penguin decays \cite{ciuchini} and
any contributions from $u$ and $c$ quarks in the mixing box diagrams, the
measured phase  is given by the following relation of CKM matrix elements:
$\phi_s^{J/\psi \phi} = 2 arg(\frac{V^*_{ts}V_{tb}}{V^*_{cs}V_{cb}}) =
-2\beta_s$.\\
The phases $\phi_s=arg(-\frac{M_{12}^s}{\Gamma_{12}^s})$ and $\phi_s^{J/\psi
  \phi}$ are two different
quantities \cite{lenz}. There is no trivial relation between
these two observables. However both are predicted to be small in the Standard
Model and most important, any sizeable New Physics contribution will affect both observables
similarly:
\begin{center}
$\phi_s = \phi_s^{SM} + \phi_s^{NP}$;~~~~~$-2\beta_s = -2\beta_s^{SM} +
  \phi_s^{NP}$;
\end{center}

\section{The $\phi_s$ Analysis Roadmap }
The decay $B_s \rightarrow J/\psi \phi$ is a pseudo-scalar to vector-vector
decay. Angular momentum conservation implies that the final state is an
admixture of $CP$-even and $CP$-odd components. By performing a time-dependent
angular analysis using the transversity angles $\Omega = (\theta, \phi, \psi)$
(Fig. \ref{fig:SHM_3}) it is possible to statistically disentangle the different $CP$ eigenstates
by the differential decay rate for $B_s$ and $\bar{B_s}$ mesons produced as
flavour eigenstates at $t=0$ is given by:
\begin{center}
$\frac{{\mathcal{\partial}^4} \Gamma(B_s \rightarrow J/\psi
    \phi)}{\mathcal{\partial}t~\partial \cos\theta~\partial \phi~\partial \cos\psi} \propto \sum_{k=1}^{6} h_k(t)f_k(\theta,\psi, \phi)$
\end{center}
The definitions of the time angular dependent functions are given in \cite{roadmap}.
$\phi_s^{J/\psi \phi}$ typically appears in these functions multiplied with terms
such as $\sin(\Delta m_s t)$. Since these terms have opposite sign between $B_s$ and
$\bar{B_s}$ the analysis benefits significantly from flavour tagging,
especially for small values of $\phi_s^{J/\psi \phi}$. We extract a measurement of $\phi_s^{J/\psi
  \phi}$ by performing an unbinned negative log-likelihood to the proper time $t$,
$\Omega$, $B_s$ mass and initial flavour tag of the selected $B_s \rightarrow
J/\psi \phi$ events. Additional physics parameters in the fit are 
$\Delta \Gamma_s$, $\Gamma_s$, the $CP$ amplitudes ($A_\perp$, $A_\parallel$, $A_0$) and
corresponding strong phases ($\delta_{\perp}, \delta_{\parallel},
\delta_0$). Parameters describing the background, and detector effects such as
the proper time resolution, mistag probability and angular acceptances
corrections are included in the PDF which will be fitted to the data as well.

\section{Monte Carlo Based Expectations}
Dedicated LHCb Monte Carlo simulations predict 117k triggered and
reconstructed
 $B_s \rightarrow J/\psi\phi$ signal candidates in 2 fb$^{-1}$ of data 
taken at $\sqrt{s}$ = 14 TeV, with a running scenario of in average one interaction per
bunch crossing. The following performance numbers are derived from a Monte
Carlo corresponding to these conditions.\\
A tagging performance of $\epsilon D^2$ = 6.2 $\pm$ 0.2\% is expected. About
60\% of the tagging power comes from so-called opposite side tagging
exploiting information related to the other B hadron in the event. 40\%
comes from a so-called same side tagger, which exploits fragmentation
properties of the $B_s$ signal candidate. The proper time
resolution is found to be $\sigma_{t}$= 38 $\pm$ 5 fs. 
\clearpage

\begin{figure}[ht]
\begin{minipage}{0.5\textwidth}
\includegraphics[height=4.2cm]{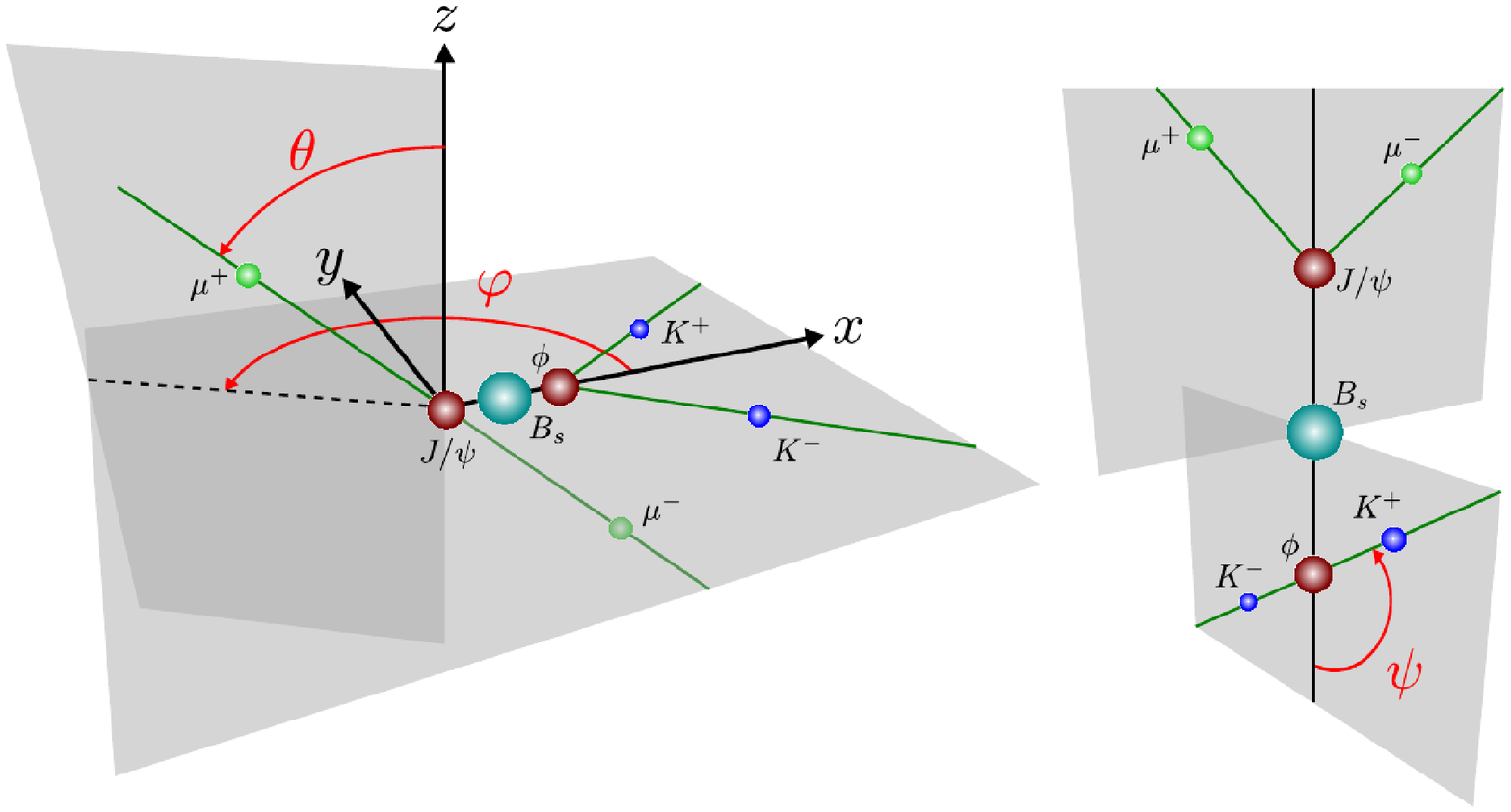}\\
\vspace{-0.2cm}

\caption{Definition of transversity base.}
\label{fig:SHM_3}
\end{minipage}
\begin{minipage}{0.465\textwidth}
\centering
\includegraphics[height=4.5cm]{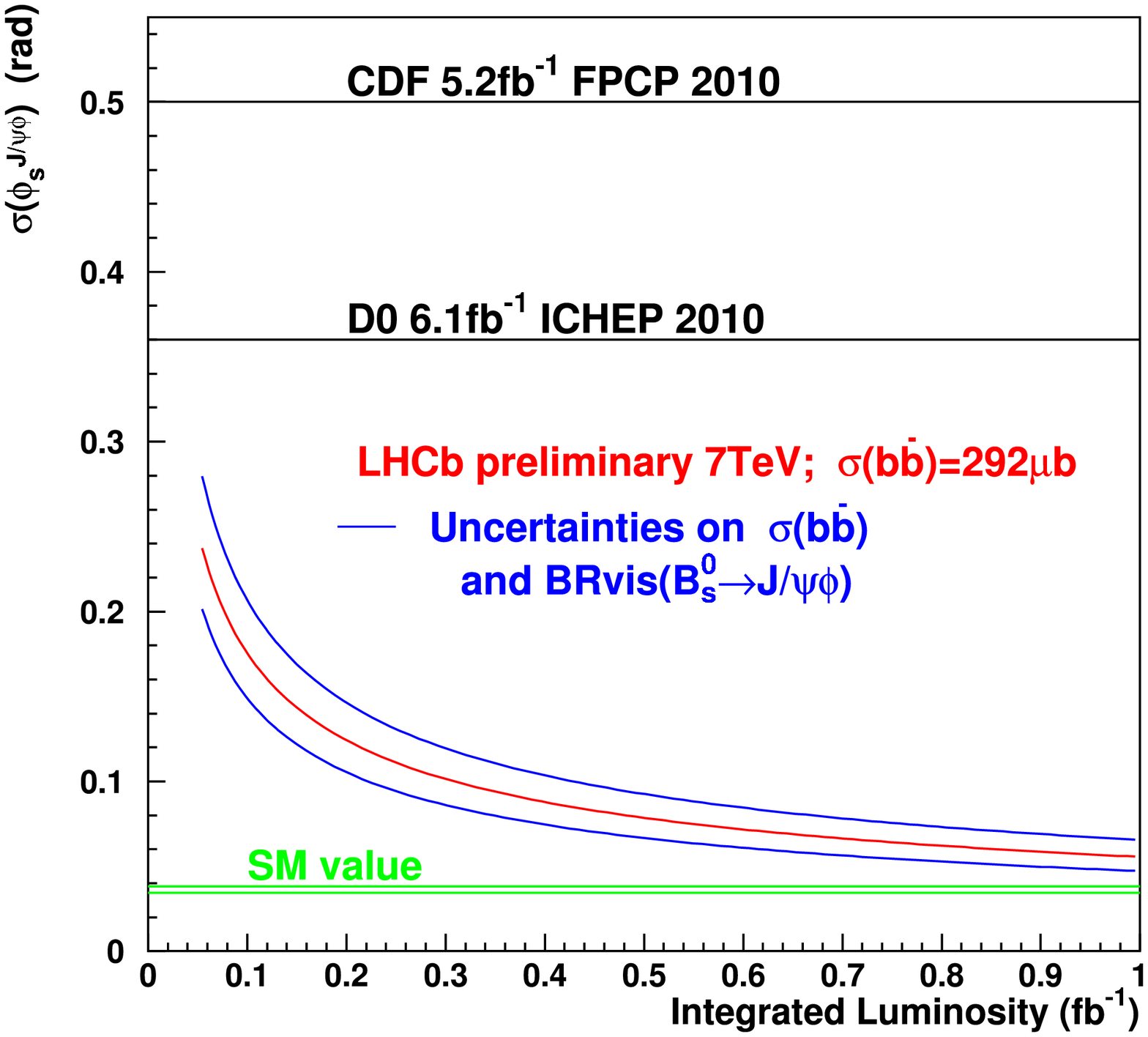}
\caption{Expected sensitivity on $\phi_s^{J/\psi\phi}$.}
\label{fig:SHM_4}
\end{minipage}
\end{figure}

The tagging performance will be calibrated on
data in various reference channels, such as $B_d \rightarrow J/\psi K^*$,
$B^+ \rightarrow J/\psi K^+$ and $B_s \rightarrow D_s \pi$. We expect a
precision from the calibration of the mistag probability $\omega$
of 3\% which results in up to 7\% relative bias on $\phi^{J/\psi \phi}_s$ for large New
Physics like values. Any bias on the proper time calibration can be
corrected in the fitter directly by according floating fit parameters, thus no
bias on $\phi^{J/\psi \phi}_s$ is expected from this source. The angular acceptances will be taken from
Monte Carlo data, however will be cross-checked in the analysis of the polarization
amplitudes of $B_d \rightarrow J/\psi K^*$ which are already known
from previous experiments \cite{jpsikstar}. Due to the limited precision of early
measurements we can validate the angular acceptances only to a accuracy of
$\pm$ 5\%. This uncertainty results in a systematic uncertainty on
$\phi^{J/\psi \phi}_s$ of up to 7\%. \\
More detailed information on the Monte Carlo studies for $\phi_s^{J/\phi
  \phi}$ can be found in \cite{roadmap}.\\
The LHC started to take data at a center-of-mass energy of $\sqrt{s}$ = 7 TeV
end of March this year. Two independent analysis using $B\rightarrow D\mu\nu X$ and displaced $J/\psi$
candidates resulted in the first measurement of the $b\bar{b}$ cross section at 7
TeV \cite{sheldon}. Combining this measurement with 
the expectations from Monte Carlo in terms of tagging
performance, proper time resolution, background levels and signal
reconstruction efficiency results in a sensitivity on $\phi_s^{J/\psi\phi}$
which is displayed in Figure \ref{fig:SHM_4}. For this study the $CP$
amplitudes and relative strong phase have been taken from recent measurements
in the analysis of $B_d \rightarrow J/\Psi K^*$ \cite{jpsikstar}. The world
average values on $\Gamma$
and $\Delta \Gamma$ have been used \cite{PDG} and $\phi_s$ was set to Standard Model
theory predictions \cite{nierste}.

\begin{figure}[htb]
\includegraphics[height=3.5cm]{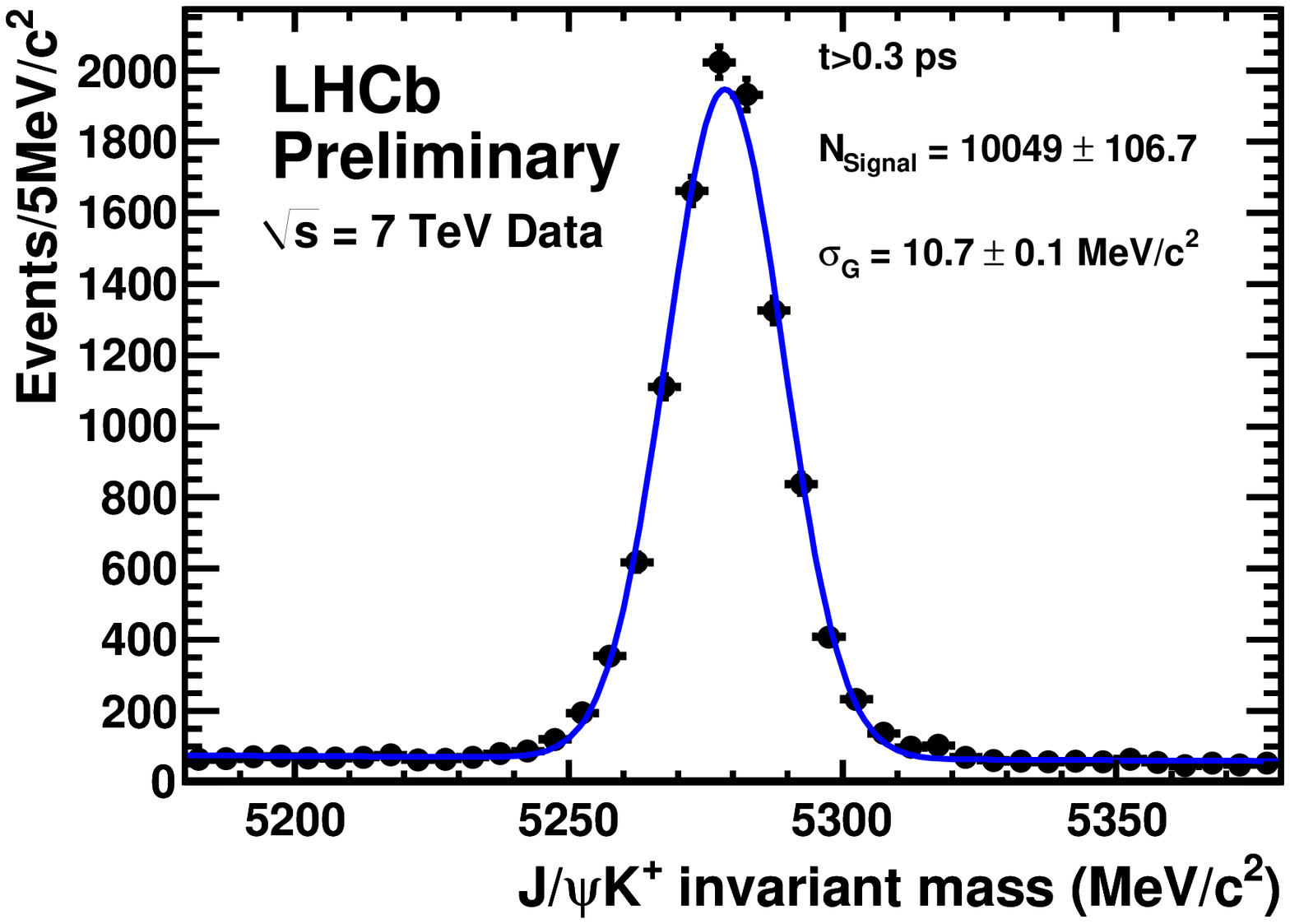}~
\includegraphics[height=3.5cm]{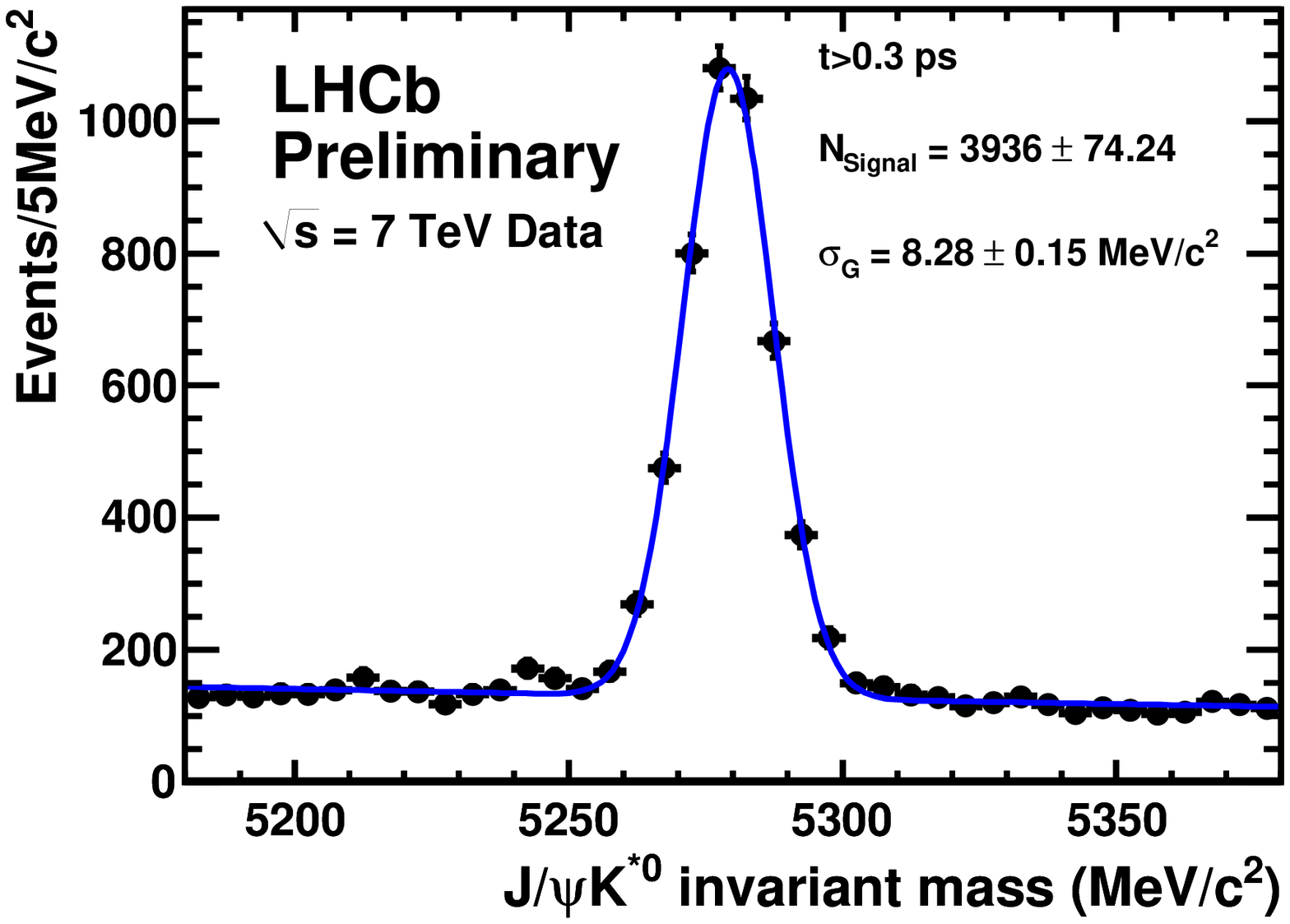}~
\includegraphics[height=3.5cm]{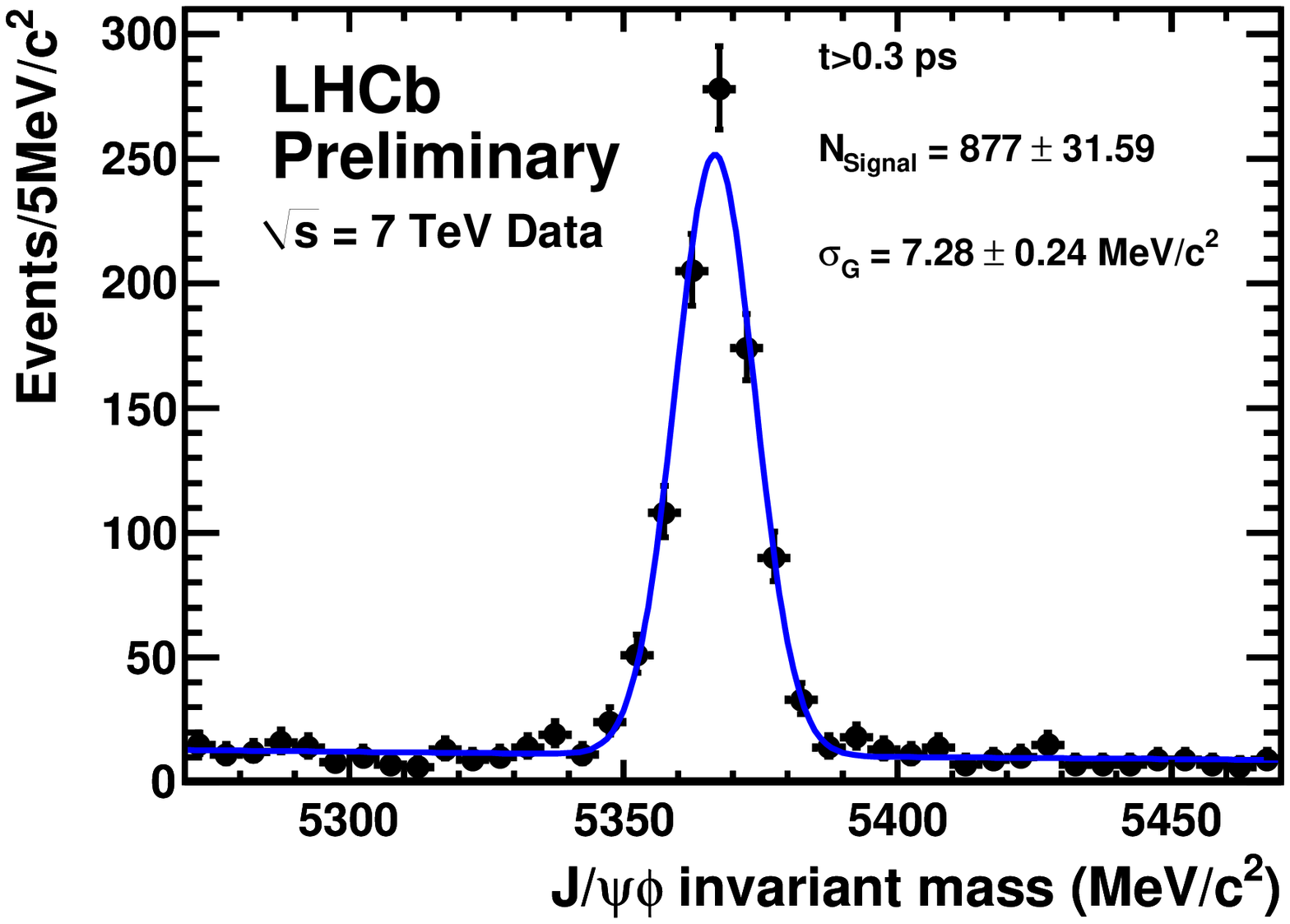}
\caption{Reconstructed $B^+ \rightarrow J/\psi K^+$, $B_d \rightarrow J/\psi
  K^*$ and $B_s \rightarrow J/\psi \phi$ candidates in a data set of 34
  pb$^{-1}$. The cut on t $>$ 0.3 ps is for illustrating purposes, but will
  not be applied in the analysis. Quoted event numbers in the plots include this cut.}
\label{fig:SHM_5}
\end{figure}

\section{First Look at Signal Candidates in Data}

At the time of the CKM workshop several $B \rightarrow J/\psi X$ signals based
on a sample of up to 3 pb$^{-1}$ have been established. Rates of 400 $B^+
\rightarrow J/\psi K^+$, 200 $B^0 \rightarrow J/\psi K^*$ and 45 $B_s
\rightarrow J/\psi \phi$ per pb$^{-1}$ of data have been found, however with
rather large uncertainties. 
A proper time resolution of about 78 fs was reported in data. No tagging
peformance was established at that time.\\
In the meantime a sample of an integrated luminosity of 37.5 pb$^{-1}$ is 
available which will be the basis of the physics results presented at 
the winter conferences 2011. In 34 pb$^{-1}$ of this sample we find a rate of 360, 150, 30,
$B^+$, $B_d$ and $B_s$ candidates per pb$^{-1}$ respectively (Fig. \ref{fig:SHM_5}). 
The reason for the non linear scaling with the luminosity is related to 
the changed trigger and running conditions. While the initial data set 
consists of events with in average one interaction per event, a good fraction 
of the full data set has been taken with in average up to three interactions per event.\\
It turned out that the extraction of the proper time resolution presented at
the CKM workshop was ignoring background contributions from fake $J/\psi$
candidates, which resulted in an overestimate of the proper time resolution. \\
The dilution on the measured mixing amplitude due to the proper time
resolution has been found to be equivalent to a Gaussian of about 50 fs.
This is about 20-30\% worse than the prediction from Monte Carlo.\\
First result of the opposite side tagging in $B_d \rightarrow
D^*\mu \nu X$ decays in data found 60\% of the expected performance. Further
improvements in this area are ahead.
No results on the same side tagger performance are available yet.

\section{Extraction of $\phi_s$ in Semileptonic Asymmetries}
An alternative way to access $\phi_s$ is to measure the final state
asymmetry $a_{fs}^s$ in flavour specific $B_s$ decays. 
This asymmetry is realted to $\phi_s$ in the following way:
\begin{center}
$a^s_{fs} = \frac{\Delta \Gamma^s}{\Delta m_s}\tan \phi_s$
\end{center}
However the measure asymmetry (both in the $B_d$ or $B_s$ system) is given by:
\begin{eqnarray}
A^q_{fs}(t) &=& \frac{\Gamma(f)-\Gamma(\bar{f})}{\Gamma(f)+\Gamma(\bar{f})} \nonumber \\
           &=& \frac{a^q_{fs}}{2} - \frac{\delta^q_c}{2} -
  (\frac{a_{fs}^q}{2}+\frac{\delta_p^q}{2})\frac{\cos(\Delta
  m_qt)}{\cosh(\Delta \Gamma_qt/2)} + \frac{\delta^q_b}{2}(\frac{B}{S})^q \nonumber
\end{eqnarray}

Several additional contributions such as the detector asymmetry $\delta_c$
($\sim 10^{-2}$), production asymmetry $\delta_p$ ($\sim 10^{-2}$) and background
asymmetry $\delta_b$ ($\sim 10^{-3}$) make the extraction of the very small value of $a_{fs}^q$ impossible.\\
However by studying simultaneously $B_s$ and $B_d$ decays with the same final
state particles such as $B_s \rightarrow D^-_s(K^+K^-\pi^-)\mu^+ \nu$ and $B_d \rightarrow
D^-(K^+K^-\pi^-)\mu^+\nu$ the detector asymmetry cancel in the difference $A^d_{fs} -
A^s_{fs}$.
Performing a time dependent analysis it is possible to extract $a^d_{fs} -
a^s_{fs}$.
Given the huge statistics expected at the LHC for next year, this analysis
will have tiny statistical uncertainties, however to control
systematic effects to the required precision will be very challenging.

\section{Summary}
The LHCb experiment successfully started data taking. First physics signals
towards the measurement of $\phi^{J/\psi \phi}_s$ have been established. The huge statistics
and very good performance of the experiment results in excellent perspectives for a
precision measurement of CP violation in the $B_s$ system with the data taken
in 2011.


{\footnotesize
\begin{thebibliography}{99}

\bibitem{ciuchini} Marco Ciuchini, ``Theoretical uncertainty on $B\rightarrow
  J/\psi K$'', same proceedings
\bibitem{nierste} See for example:\\
U. Nierste, ``Bounds on New Physics from $B_s^0$ mixing'',
Int. J. Mod. Phys. 22, 5986 (2008);\\
A. Lenz, ``Unparticle physics effects in $B_s^0$ mixing'', Phys. Rev. D 76,
065006 (2007);\\
R. Fleischer, ``CP Violation and B Physics at the LHC'',
arXiv:hep-ph/0703112v2 (2007).
\bibitem{sheldon} The LHCb Collaboration, ``Measurement of $\sigma(pp
  \rightarrow b\bar{b}X)$ at $\sqrt{s}$=7TeV in the forward region'',  Phys. Lett. B 694 (2010) 209.
\bibitem{angularanalysis} 
A. Dighe {\itshape et al.,} ``Angular distribution and lifetime difference in
$B_s^0 \rightarrow J/\psi \phi$ decays'', Phys. Lett. B 369, 144 (1996).\\
K. Anikeev {\itshape et al.,} ``B Physics at the Tevatron: Run II and beyond'',
arXiv:hep-ph:0201071v2 (2002).
\bibitem{lenz} A. Lenz, ``Theoretical status of $B_s^0$-mixing and lifetimes of
heavy hadrons'', arXiv:0705.3802v2 [hep-ph] (2007)
\bibitem{roadmap} The LHCb Collaboration, ``Roadmap for selected key
  measurements of LHCb'', arXiv:0912.4179v2 [hep-ex]
\bibitem{jpsikstar} Babar Collaboration, ``Measurement of decay amplitudes of
  $B \rightarrow J/\psi K^*, \psi(2S)K^*$, and $χ_{c1}K^*$ with an angular analysis'', Phys. Rev. D, 76, 031102 (2007)\\
                    Belle Collaboration, ``Studies of CP violation in
                    $B \rightarrow J/\psi K^*$ decays'', Phys. Rev. Lett. 95, 091601 (2005)
\bibitem{PDG} C. Amsler {\itshape et al.} (Particle Data Group), Phys. Lett. B
  667, 1 (2008)
\bibitem{asf} N. Brook {\itshape et al}, "LHCb's potential to measure
  flavour-specific CP-asymmetry in semileptonic and hadronic decays'', LHCb
note 2007-054 (2007)

\end{thebibliography}
}
\end{document}